\documentclass[aps,twocolumn,groupedaddress,longbibliography]{revtex4-1}

\usepackage{graphicx}
\usepackage{dcolumn}
\usepackage{bm}
\usepackage{bbm}
\usepackage{amsfonts}
\DeclareMathAlphabet\mathbfcal{OMS}{cmsy}{b}{n}
\usepackage{amsmath}

\newcommand\redout{\bgroup\markoverwith{\textcolor{red}{\rule[.5ex]{2pt}{0.4pt}}}\ULon}

\usepackage[colorlinks=true,citecolor=blue]{hyperref}
\hypersetup{colorlinks=true,citecolor=blue,linkcolor=red,urlcolor=blue}

\begin{document}

\title{Topological states of generalized dissipative Majorana wires}

\author{Farokhnaz Hosseinifar}
\affiliation{Department of Physics, Institute for Advanced Studies in Basic Sciences (IASBS), Zanjan 45137-66731, Iran}
\affiliation{Research Center for Basic Sciences \& Modern Technologies (RBST), Institute for Advanced Studies in Basic Science (IASBS), Zanjan 45137-66731, Iran}

\author{Ali G. Moghaddam}
\affiliation{Department of Physics, Institute for Advanced Studies in Basic Sciences (IASBS), Zanjan 45137-66731, Iran}
\affiliation{Research Center for Basic Sciences \& Modern Technologies (RBST), Institute for Advanced Studies in Basic Science (IASBS), Zanjan 45137-66731, Iran}

\date{25 July 2022}

\begin{abstract}
We study the generalized one-dimensional (1D) quantum dissipative models corresponding to a Majorana wire which can possess more than one Majorana bound state at each end. The system consists of a 1D fermionic open quantum system whose dynamics is governed by a quadratic Lindblad equation. Using the adjoint Lindblad equation for the fermionic two-point correlations, we find the gaps in the damping and purity spectra of a generic 1D model. Then, using the symmetry-based classification, we show that a winding number as the topological invariant can be defined which distinguishes different steady states of the system in the presence of damping and purity gaps. Then we focus on certain models with different Lindblad quantum jump terms and explore their phase diagrams by calculating the damping and the purity gaps as well as the winding number. In particular, we show that by inclusion of quantum jumps between next-nearest-neighbor sites, higher winding numbers and equivalently more Majorana bound states can be achieved. Also, by introducing imbalanced couplings we can switch between states with negative and positive winding numbers. Finally, we should mention that since our formulation is based on the fermionic correlations rather than the Majorana operators, it can be easily extended to the dissipative topological phases belonging to other symmetry classes.
\end{abstract}

\maketitle

\section{Introduction}
Over the past two decades, predictions and discoveries of new topological phases have opened a new era in condensed matter physics, if not physics as a whole \cite{Kane2010RMP,Zhang2011RMP}. Besides the interest in the fundamental aspects of these new phases of matter, they have been considered very promising in various applications in electronics, spintronics, photonics, and even quantum technologies \cite{Ozawa2019,Nayak2008}. Most of these potential applications rely on the cornerstone of topological phases as their protection against perturbations, disorder, and imperfections besides the bulk-boundary correspondence that guarantees the existence of low-energy boundary modes opposed as to the gapped bulk states \cite{Hatsugai,graf2013bulk,prodan2016bulk}. A key insight into the topological phases has been developed by considering the role of  symmetries in topological protection which has led to the tenfold classification of topological states, also known as the periodic table of topological insulators and superconductors \cite{Kitaev2009,Ryu2010,Ryu2016}. 

An important question that has remained to answer was if and how the topological properties maintain under non-equilibrium situations mainly in the presence of dissipation and external driving.
In such situations, the dynamics of quantum systems becomes more involved and complicated than pure unitary evaluations governed solely by a Hamiltonian.
Strikingly, it has been found that not only topological properties can persist in non-equilibrium systems 
\cite{Jiang2011}, but starting from non-topological equilibrium states, a nontrivial topological phase can be obtained by dissipation and driving \cite{diehl2011topology,Bardyn2013,gefen-DissipativeMajorana,Castillo,Budich2015-dissi}. The basic idea is to consider systems with a certain form of dissipative dynamics described by the quantum master equation, which leads to topologically nontrivial steady states.
To characterize the topology of mixed quantum states represented by density operators, 
various methods have been suggested, among them one mathematically elegant but practically elusive approach is based on the formal generalization of the geometric phases known as the Uhlmann phase \cite{uhlmann1986parallel,Martin-Delgado2014,Arovas2014,Budich2015}.
More recent studies have shown that topological features of mixed Gaussian quantum states in one dimension can be probed through the so-called
ensemble geometric phase \cite{Bardyn-PRX2018,Fleischhauer2020,Fleischhauer}.
Subsequently, topological states in far-from-equilibrium fermionic open quantum systems 
have been fully classified based on the symmetries of the dynamical equation governing the density matrix evolution, in particular those of quadratic Lindbladians \cite{Cooper2020,Cooper2019,Altland2021}.

In order to achieve a dissipative topological phase, the dissipative dynamics have usually been constructed such that the system relaxes to a decoherence-free dark state which corresponds to a ground state of a nontrivial topological phase \cite{diehl2008quantum,MULLER2012,Diehl-TQFT-nonequilibrium}.  
One particular example of such models that has been intensely studied is the non-equilibrium corresponding of the 1D Kitaev chain that hosts isolated Majorana bound states in their long-term steady behavior \cite{Bardyn2013,diehl2011topology,gefen-DissipativeMajorana,Viola2021}. Although various aspects of the dissipative version of Kitaev chain has been explored extensively, its generalization to models with higher topological numbers which can host more Majorana bound states has been overlooked so far \cite{Castillo}. Here, we introduce generalizations of the dissipative Kitaev model with higher topological numbers and obtain the phase diagram of topological states for particular examples of such generalized models. In these models, we consider imbalanced dissipative couplings of the fermionic sites to the bath as well as additional couplings beyond nearest neighbors. 
We find that for the model with imbalanced couplings, the relative sign of excitation and de-excitation terms can change the topological phase of the system. In the model with next-nearest-neighbor (n.n.n.) couplings, we show that a minimum strength of such additional terms is required to achieve a phase with a higher topological number.

\section{Dissipative lattice models} 

\begin{figure}
\centering
\includegraphics[width=0.8\linewidth]{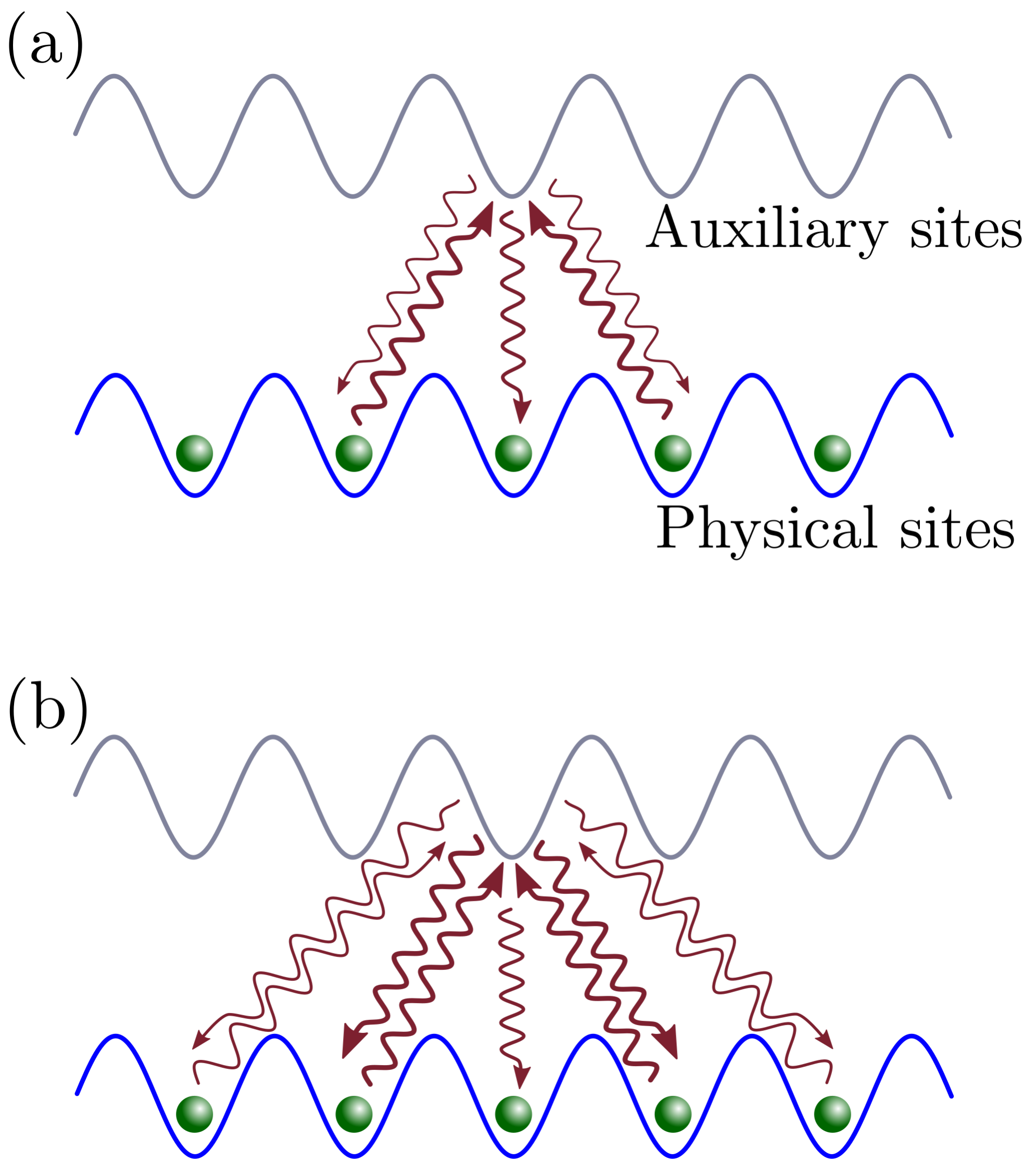}
\caption{1D cold atom chain in an optical lattice. The physical sites in the lower parts of each panel are coherently coupled to the auxiliary sites driven by external lasers. There is also dissipation via spontaneous emission, which brings the states back to the lower sites. Driving and dissipation are represented by curvy arrows inclined upwards and downwards, respectively. 
(a) Imbalanced coupling model with nearest-neighbor (n.n.) sites involved. Coherent driving couples atoms at each pair of n.n.n. sites to the auxiliary state corresponding to the site between them. The relaxation the from auxiliary state though takes place towards the corresponding physical site and its nearest neighbors. The excitation and relaxation (de-excitation) rates between n.n. sites are assumed to be different, as indicated by thicker and thinner arrows, respectively.
(b) Shows a similar system but without imbalance but with additional couplings between n.n.n. sites to the auxiliary sites.  
}
\label{fig:lattice}
\end{figure}

The time evolution of the reduced density matrix of a system in the presence of an external bath follows the Liouville equation that under Born-Markov approximations reduces to the Lindblad equation \cite{lindblad1976,gorini1976,rivas2012open,Rudner2020}
\begin{eqnarray}
\partial_t \rho = -i [H, \rho] + \sum_i \left( l_i \rho l_i^{\dagger} - \frac{1}{2}\lbrace l_i^{\dagger}l_i, \rho \rbrace \right).
\end{eqnarray}
The first term on the right-hand side corresponds to the coherent unitary dynamics under the 
system Hamiltonian, whereas the second term, called \emph{Liouvillian}
or \emph{Lindbladian}, represents the dissipative dynamics due to the coupling to the bath. Following the pioneering works
on topology in dissipative systems
\cite{diehl2011topology,Bardyn2013},
we concentrate on purely dissipative dynamics and set $H=0$, provided by large energy splitting $\Delta$
between
the lowest and excited states of the system such that we can truncate the system Hilbert space to its ground state only. 
The Lindblad operators
$l_{i}$ represent the quantum jumps in the system evolution.
We further concentrate
on a driven-dissipation dynamics
with quasi-local excitation/de-excitation 
processes for which the Lindblad operators can be written in terms of particle creation and annihilation operators as
\begin{align}
    l_{i} = \sum_{j,j'~{\rm around }~i} \ell_{jj'}\: a_j^\dag a_{j'}, \label{eq:lindblads-1}
\end{align}
where the summation runs over a certain neighborhood of the physical site $i$.
The coefficients $\ell_{jj'}$
are determined by the details of coupling between the system and the bath through external driving
and dissipation processes.
Such dynamics can be realized in cold atom settings by applying spatially modulated laser fields giving rise to  off-resonance Rabi frequencies $\Omega\ll\Delta$ as extensively studied in previous works \cite{diehl2008quantum,MULLER2012}. 
The underlying physical mechanism for quantum jumps represented by Eq. \eqref{eq:lindblads-1}
consists of two steps: (a) driven excitation to auxiliary states due to the Laser field and (b) relaxation back to the low-lying
physical states due to the dissipation (for instance, through spontaneous emission of phonons to the bath). Such driven-dissipative dynamics of cold atoms in an optical lattice are schematically illustrated in Fig. \ref{fig:lattice}.

Now the Lindblad equation 
with jump operators \eqref{eq:lindblads-1}
represents a quartic or interacting dynamics. Fortunately, as shown in Ref. \cite{Bardyn2013},
we can safely switch from fixed-number to fixed-phase ensembles in the thermodynamic limit
and then apply a mean-field description that eventually enables us to replace the original quadratic particle-conserving Lindblad operators with the phase-conserving linear operators 
\begin{align}
l_{i} ~\to~   L_{i} = \sum_{j~{\rm around }~i} \Big( \tilde{\ell}^{(+)}_j a_j^\dag + 
    \tilde{\ell}^{(-)}_j
    a_{j}\Big),
\end{align}
in analogy with the BCS theory of superconductivity \cite{Bardyn2013}. 
Assuming the periodic boundary condition (PBC), we can find the corresponding Lindbald operators in $k$-space which read
\begin{eqnarray} \label{eq:L-k}
L_k = \frac{1}{N}\sum_n e^{i k n}L_n = v_{k} a_k^\dag - u_k a_{-k},
\end{eqnarray}
with coefficients $v_k$ and $u_{k}$ determined from $\tilde{\ell}^{(\pm)}_j$. 
So the Lindblad equation under the assumption of $H=0$ reduces to \cite{Prosen_2008,prosen2010spectral,diehl2011topology,Bardyn2013,Cooper2019}
\begin{align}
\label{lindblad-k}
\partial_t \rho=
   \sum_{k} \Big( L_k  \rho  L_k^\dag - \frac{1}{2}\{ L_k^\dag L_k, \rho \}\Big).
\end{align}
This Lindblad equation with jump operators 
$L_k$ (or equivalently $L_{i}$ in real space) is quadratic in the fermion operators representation. Consequently,
we have free-field-type dissipative dynamics whose
solutions take the form of Gaussian density operators
\begin{align}
    \rho(t) & = \frac{\exp\Big(\sum_{nm} 
    \hat{\psi}_{n}^\dag \: \hat{h}_{nm}(t) \: \hat{\psi}_{m}
    \Big)}{\cal Z},
    \label{entH-real}
    \\ 
    & = \frac{\exp\Big(\sum_{k} 
    \hat{\Psi}_{k}^\dag \: \hat{H}_{k}(t) \: \hat{\Psi}_{k}
    \Big)}{\cal Z},
    \label{entH-k}
\end{align}
in real and momentum spaces, respectively.
In above expressions, we have $\hat{\psi}_{n}^\dag = (a_{n}^\dag,a_n)$, $\hat{\Psi}_{k}^\dag = (
a_{k}^\dag,a_{-k})$, and corresponding relations hold for their Hermitian conjugates. Equations \eqref{entH-real} and \eqref{entH-k}, indeed, define the entanglement Hamiltonian ${\cal H}_{E}$ 
associated with the density matrix through $\rho\propto \exp(-{\cal H}_{E})$. 
The key underlying reason for having Gaussian density matrices is that the entanglement Hamiltonian of the free fermions is also of a free-fermion form \cite{eisler2017analytical,peschel2003calculation}.

\section{Correlation matrix and topology}
It has been shown that all information of 
Gaussian density matrices is encoded in the  correlation matrix $\mathbfcal{C}$, whose elements consist of
expectation values of $a_n^\dag a_m$ and $a_n^\dag a_m^\dag$ terms in real space \cite{peschel2003calculation,peschel2004reduced,peschel2009reduced}.
In $k$-space, and in the presence of translational symmetry, $\mathbfcal{C}$ takes a block-diagonal form in $k$-space
with blocks 
\begin{equation}
\hat{\cal C}_k=
    \begin{pmatrix}
    \langle 
    a_{k}^\dag a_{k} \rangle
    &
    \langle a_{-k}  a_{k}  \rangle\\
    \langle a_{k}^\dag a_{-k}^\dag \rangle
    &
     \langle  a_{-k} a_{-k}^\dag
    \rangle
    \end{pmatrix} .
\end{equation}
We immediately realize that the correlation matrix elements in the above expression can be interpreted as the spectral sum of the Nambu (superconducting) Green's function
\begin{align}
\hat{\cal C}_k\equiv  \hat{\cal G}(\Delta t=0 ,k) =\int d\omega\: \hat{\cal G}(\omega ,k),    
\end{align}
with
\begin{align}
    \hat{\cal  G}(\omega ,k) =\begin{pmatrix}
    G(\omega ,k) & F(\omega ,k) \\
    F^\ast(\omega ,k) &  -G(\omega ,k)
    \end{pmatrix},
\end{align}
consisting of normal and anomalous parts of the Nambu Green's function ($G$ and $F$, respectively).
We should mention that in previous studies \cite{Bardyn2013,diehl2011topology}, the correlation matrix has been formulated in a slightly different manner and in terms of the correlations
$\langle c_{m}c_{n} \rangle$ between Majorana operators 
$c_{2n}=a_n+a_n^\dag$ and $c_{2n-1}=i(a_n-a_n^\dag)$. But one can immediately notice that the two formulations using fermionic or Majorana operators, respectively, are equivalent to each other, thus we can use either of them depending on our purpose.

\subsection{Dissipative dynamics and associated spectra}
Provided by the equivalence relation between the Gaussian density matrices and the corresponding correlations, we only need to obtain the correlation matrix elements. Therefore, instead of Eq. \eqref{lindblad-k},
we can concentrate
on the \emph{adjoint Lindblad equation} 
\begin{align}
    \label{lindblad-adjoint}
\partial_t {\cal O}=
   \sum_{k} \Big( L_k^\dag  {\cal O}  L_k - \frac{1}{2}\{ L_k^\dag L_k, {\cal O} \}\Big),
\end{align}
for any Heisenberg operator ${\cal O}(t)$.
In particular, we are interested in the two-point correlation operators
$a_{k}^\dag a_{k}$, $a_{-k} a_{k}$, $a_{k}^\dag a_{-k}^\dag$, and $a_{-k} a_{-k}^\dag$
whose quantum-statistical expectation values
yield the correlation matrix block ${\cal C}_k$.
Exploiting the fermionic anti-commutation relations and after some algebra (see Appendix \ref{app-a} for more details), the time evolution of the correlation operators reads
\begin{align} \label{key-dynamics}
    \partial_t{\cal A}_{k} = -{\cal X}_{k} {\cal A}_{k}+
    {\cal Y}_{k},
\end{align}
where 
\begin{align}\label{key-dynamics-details}
\begin{array}{cc}
     &     {\cal A}_{k}=\begin{pmatrix}
    a_{k}^\dag a_{k}\\
    a_{-k} a_{-k}^\dag\\
    a_{-k} a_{k}\\
    a_{k}^\dag a_{-k}^\dag)
    \end{pmatrix},\quad
    {\cal Y}_{k}=\begin{pmatrix}
    |v_k|^2\\
    |u_k|^2\\
    - v_k u_k^\ast\\
    -u_k v_k^\ast
    \end{pmatrix}, \\
    & \\
     & 
    {\cal X}_{k}=
     \big( |u_k|^2 + |v_k|^2 \big) 
     \hat{\tau}_0 \otimes
     \hat{\sigma}_0
      \\
    &\qquad - \,\Re(u_k v_k^\ast)\hat{\tau}_x \otimes(\hat{\sigma}_0+\hat{\sigma}_x).
\end{array}
\end{align}
The matrix ${\cal X}_k$ governs the dissipative dynamics of the system, and its eigenvalues 
determine the relaxation rates of different modes. By direct inspection, we find
four eigenvalues $\omega_k^{(1,2)}=|u_k|^2+|v_k|^2$, and $\omega_k^{(3,4)}=|v_k\pm u_k^\ast|^2\equiv \omega^{(1)} \pm 2\Re(v_k u_k^\ast)$. Subsequently, the gap in the dissipation spectrum known as the \emph{dissipative} or \emph{damping gap}
is given by $\Delta_d = {\rm min}\:\omega_k^{(3,4)}  = {\rm min}\big(|v_k\pm u_k^\ast|^2\big)$ when we sweep over all $k$'s.

In the long-term steady-state limit, $\partial_t \rho =0 $, and the expectation values become stationary, thus we have $\langle {\cal A}_k\rangle_{\rm st} = {\cal X}_{k}^{-1} {\cal Y}_{k} $, which finally gives rise to the correlation matrix
\begin{align}
    \hat{\cal C}_k &= \frac{1}{|u_k|^2 + |v_k|^2 }\begin{pmatrix}
    |v_k|^2 & -i\Im(v_k u_k^\ast)\\
   i\Im(v_k u_k^\ast) & |u_k|^2 \\
    \end{pmatrix} .
\end{align}
The above form can be recast as
\begin{equation}\label{eq:C-matrix-d-vector}
    \hat{\cal C}_k = \frac{1}{2}\hat{\sigma}_0 +\frac{1}{2} {\bf d}_{k}\cdot {\bm\sigma},
\end{equation}
which is expressed entirely in terms of a ${\bf d}$-vector:
\begin{equation}
    {\bf d}_{k}= \frac{1}{|u_k|^2 + |v_k|^2}\big[0,2\,\Im(v_k u_k^\ast), |v_k|^2 - |u_k|^2\big]. 
\end{equation}
Likewise, the correlation matrix possesses a dispersion relation
\begin{equation}
    \xi_k = | {\bf d}_{k}|=\sqrt{1-\bigg[\frac{2\Re(u_k v_k^\ast)}{|u_k|^2 + |v_k|^2}\bigg]^2},
\end{equation}
which we can refer to it as the \emph{purity spectrum} since it directly relates to the purity of the states and also the population of each stationary eigenmode as the following. Firstly, we see that this function is
limited to $0\leq \xi_k\leq 1$. Secondly, the eigenvalues of the correlation matrix are given by $(1\pm\xi_k)/2$,
which means that for $\xi_k=1$ they become $0$ and $1$.
But only for pure states in which the corresponding eigenmode is either fully occupied or empty, the correlation matrix eigenvalues are either $0$ or $1$. Consequently, the purity dispersion of a pure Gaussian state becomes flat and identically equal to $1$ for all $k$'s. On the other hand,
as long as $\xi_k>0$ for all $k$'s, 
the correlation matrix spectrum retains a gapped spectrum around $1/2$ with a \emph{purity gap} given by $\Delta_{p}={\rm min}(\xi_k)$.

\subsection{Topological invariant}
The correlation matrix expressed by Eq. \eqref{eq:C-matrix-d-vector}
provides a natural playground to study
the topological properties of the corresponding Gaussian states, similarly to Dirac Hamiltonians $\hat{\cal H}_D (k)= {\bf d}_{k}\cdot {\bm \sigma}$.
In fact, not only the topological aspects of a Gaussian state and the corresponding Dirac Hamiltonian are the same, they also share similar symmetries.
This can be understood from the fact that the correlation matrix $\hat{\cal C}_{k}$ and ${\cal H}_D$ only differ in a term proportional to the identity matrix which is insensitive to any symmetry operation or topology. We note that the symmetries of the steady-state correlation matrix
are always inherited from those of the Lindblad operators \cite{Bardyn2013}.
\par
To study the topological properties of the Dirac Hamiltonian,
we can take advantage of the periodic table of topological phases, which is based on symmetries \cite{Ryu2016,Bardyn2013}.
We can readily see the symmetry relation 
\begin{equation}\label{TRS}
    \hat{\cal H}_D^\ast(-k) = \hat{\cal H}_D(k),
\end{equation}
since $d_y(-k) = \Im(v_{-k} u_{-k}^\ast) = - \Im(v_{k} u_{k}^\ast)=-d_y(k)$, $d_z(-k) =   |v_{-k}|^2 -|u_{-k}|^2 =  |v_{k}|^2 -|u_{k}|^2 = d_z(k)$ \footnote{We should also remember that $\sigma_y^\ast=-\sigma_y$ while $\sigma_z^\ast=\sigma_z$}.
This relation corresponds to the time-reversal symmetry $\hat{T} \hat{\cal H}_{-k} \hat{T}^{-1} = \hat{\cal H}_{k}$ by $\hat{T} \equiv {\cal K}$ being just a complex conjugation operator. 
Another symmetry relation directly follows the fact that
the Dirac Hamiltonians corresponding to our model do not include the Pauli matrix
$\hat\sigma_x$. Therefore, they satisfy the relation
\begin{equation}\label{Chiral}
    \hat\sigma_x \: \hat{\cal H}_D (k) \:\hat\sigma_x  = -\hat{\cal H}_D(k),
\end{equation}
due to the basic anticommutation properties of Pauli matrices, $\{\hat\sigma_i,\hat\sigma_j\}=\delta_{ij}$.
Equation \eqref{Chiral} indicates the chiral symmetry $\hat{S} \hat{\cal H}_{k} \hat{S}^{-1} = - \hat{\cal H}_{k}$, equivalent to an anticommutation of the Hamiltonian with the unitary operator $\hat{S} = \sigma_x$.
The simultaneous presence of the time-reversal and chiral symmetries implies
that the particle-hole symmetry $\hat{C} \hat{\cal H}_{-k} \hat{C}^{-1} = -\hat{\cal H}_{k}$
such that $\hat{S} = \hat{C} \hat{T}$ should exist.
It can be directly illustrated by combining the two symmetry relations in Eqs. \eqref{TRS} and \eqref{Chiral} that give rise to
\begin{equation}
        \hat\sigma_x \: \hat{\cal H}_D^\ast (-k) \:\hat\sigma_x  = -\hat{\cal H}_D(k),
\end{equation}
which means that the particle-hole symmetry operator is given by $\hat{C} = \hat\sigma_x {\cal K}$.
\par
Symmetry properties found above indicate that our models belong to the class BDI
of the topological phases since both time-reversal and particle-hole operators 
square to the identity operator ($\hat{T}^2=\hat{C}^2={\mathbbm 1}$).
According to the periodic table of the topological phases, 
topologically distinct 1D models in the symmetry class BDI are characterized by a  ${\mathbbm Z}$ invariant \cite{Kitaev2009, Ryu2010, Ryu2016}.
The presence of chiral symmetry in any odd dimension guarantees that we can define
a topological winding number $\nu$. 
The crucial point is that any Hamiltonian
with chiral symmetry can be brought to a purely block-off-diagonal form
(the corresponding basis is usually known as the chiral basis). 
For the models considered here, by performing a unitary transformation 
$ \hat{\cal H}'_k= \hat{U} \hat{\cal H}_k \hat{U}^\dag$ with $\hat{U}= \exp(-i\hat\sigma_y\pi/4)$, the Hamiltonian takes the desired form
\begin{equation} \label{eq:chiral-Hamiltonian}
    \hat{\cal H}'_k  
    =
    \begin{pmatrix}
        0 &  d_k^z-i d_k^y \\
        d_k^z+i d_k^y & 0
    \end{pmatrix}
    = |{\bf d}_{k}|
   \begin{pmatrix}
        0 &  q_k \\
        q_k^\ast & 0
    \end{pmatrix},
\end{equation}
where $q_k \equiv e^{i\vartheta_{k}}$
is a phase factor depending on the unit vector $\hat{\bf d}_{k}={\bf d}_{k}/|{\bf d}_{k}|$ \footnote{We should remind that, in a more general situation and for a $2n$-band Hamiltonian, $q_k$ is a $n\times n$ unitary matrix}.
We can thus define a winding associated with the phase $\vartheta_{k}$ 
as 
\begin{equation}
\nu_{\rm 1D}=\int \frac{dk}{2\pi} \: \frac{d}{dk}\vartheta_{k}=\int \frac{dk}{2\pi i} \:  q_k^\ast  \frac{d}{dk} q_{k}.
\end{equation}
Equivalently, the winding number
can be written in terms of the unit vector $\hat{\bf d}_{k}$ as
\begin{align}\label{eq:winding}
  \nu_{\rm 1D} =  \frac{1}{2\pi}\int dk \: \hat{\bf x}\cdot \Big(\hat{\bf d}_k\times \frac{d}{dk}\hat{\bf d}_k  \Big),
\end{align}
using the connection of $q_k$ and the two nonzero components of the unit vector $\hat{\bf d}_{k}$.
Given that 
the phase $\vartheta_k$ and equivalently unit vector $\hat{\bf d}_{k}$
here define maps from the 1D Brillouin zone ($T^1$) into a unit circle ($S^1$),
both relations above correspond to a well-defined 
${\mathbbm Z}$ topological invariant.
\par
In odd dimensions, it is possible to define another invariant known as the Chern-Simons (CS) invariant, along with the winding number. For 1D models, it is given
integral of Berry's connection ${\cal A}_{k}$ as \cite{Ryu2016}
\begin{equation}
    {\rm CS}_{\rm 1D} = \int \frac{dk}{2\pi i} \: {\rm Tr} {\cal A}_{k}.
\end{equation}
It should be noted that this invariant is not gauge-invariant nor quantized in general. However, in the presence of chiral symmetry and choosing a specific gauge corresponding to the chiral basis, it shows a one-to-one correspondence to the winding number. We can illustrate this by noting that the eigenstates of Hamiltonian in Eq. \eqref{eq:chiral-Hamiltonian} are 
\begin{equation}
    |u_{k,\pm}\rangle = \frac{1}{\sqrt{2}}\begin{pmatrix}
        1 \\ \mp e^{i\vartheta_{k}}
    \end{pmatrix},
\end{equation}
which results in Berry's connection ${\cal A}_{k} =\langle u_{k,-}|  d/dk  |u_{k,-}\rangle = (i/2) d\vartheta_k/dk$ and thereby a CS invariant ${\rm CS}_{\rm 1D}  = \nu_{\rm 1D}/2$. Hence, in the remainder of the paper,
we only show the winding number as the topological invariant, which in general, can take any integer value.
The topological invariants remain well-defined as long
there exists a gap such that the vector ${\bf d}_{k}$ remains nonzero throughout the Brillouin zone ($0\leq k < 2\pi$).  
Accordingly, as shown in Ref. \cite{Bardyn2013}, the topological character of the steady-states only changes if at least one of the two characteristic gaps (purity or dissipation) closes and reopens.

\section{Dissipative Kitaev models}

We start with a brief review of the basic dissipative Kitav model that has been previously studied \cite{Bardyn2013,diehl2011topology}. This model is represented by the Lindblad operator 
\begin{align}
L_n = \frac{1}{2} \Big(
a^{\dagger}_{n+1} + a^{\dagger}_{n-1}  +  a_{n+1} - a_{n-1}
\Big).   \label{eq:lindblad-model-kitaev}
\end{align} 
It has been shown that the steady state of the corresponding dynamics
becomes a pure, Dark state with a nontrivial topology that hosts two Majorana zero modes at each end of the chain in an open geometry. The Majorana end states represent a nonlocal decoherence-free subspace
that is isolated from the dissipative bulk modes by a finite dissipative gap.
Considering the periodic boundary conditions, the Fourier components of the Lindblad operator are found as $v_k=\cos k$ and $u_k=-i\sin k$, which results in a ${\bf d}$-vector
\begin{align}
    {\bf d} = -\sin 2k \:\hat{\bf y} +  \cos  2k \:\hat{\bf z}.
\end{align}
Then, using the winding number definition, Eq. \eqref{eq:winding}, we find a nonvanishing topological invariant of $\nu=2$ as a manifestation of the boundary-bulk correspondence. 

In the following, we will consider two extensions of this basic model and obtain their topological states phase diagram as well as their purity and dissipative spectra. As we will see, both models have a totally pure steady state, and their purity gap maintains throughout the parameters space. However, their dissipative gap varies with the dissipation parameters encoded in the Lindbladian. All the associated gap closings correspond to topological phase transitions where the winding number changes.

\subsection{Model I: Imbalanced couplings}
\begin{figure}[t]
\centering
\includegraphics[width=0.99\linewidth]{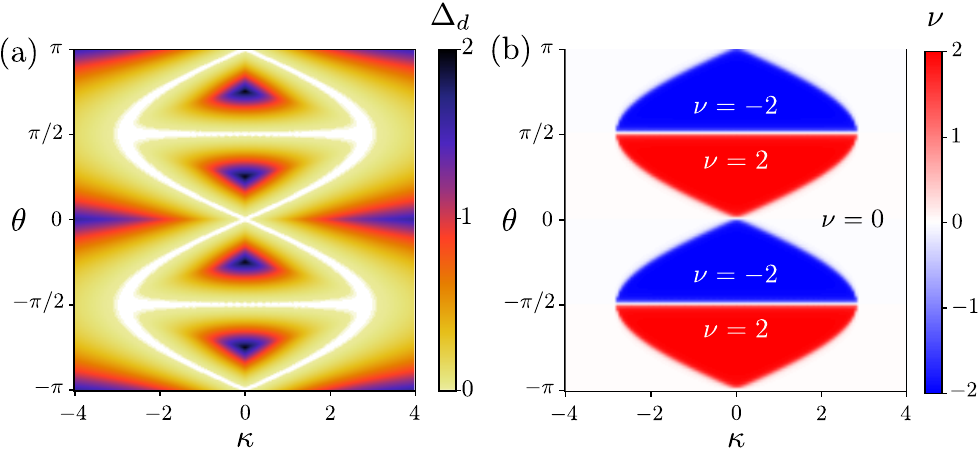}
\caption{(a) Dissipative (damping) gap of the Kitaev model with imbalanced coupling as a function of the imbalance control parameter $\theta$ and the onsite dissipation parameter $\kappa$. (b) The topological winding number variation with the two parameters $\theta$ and $\kappa$. As the model has always a nonvanishing purity gap, the change in the winding number exactly coincides with the dissipative gap closing lines. The switching between $\nu=\pm 2$ has been made possible by the imbalanced coupling controlled by $\theta$. }
\label{fig:modelI}
\end{figure}

The first model that we consider is described by the Lindblad operator
\begin{align}
 L^{\rm I}_n = \frac{1}{\sqrt{4+{\kappa}^2}} \Big[
{\kappa}\,a^{\dagger}_n &+ \sqrt{2}\sin\theta(a^{\dagger}_{n+1} + a^{\dagger}_{n-1}) \nonumber \\
&
 + \sqrt{2}\cos\theta(a_{n+1} - a_{n-1})
\Big]   \label{eq:lindblad-model-I},
\end{align} 
where $\kappa$ is a real parameter controlling the coupling to the central site $n$,
and we have asymmetric couplings 
(with imbalanced excitation/de-excitation rates)
to the neighboring sites as illustrated in Fig. \ref{fig:lattice}(a).
The Fourier components of the Lindblad operator in $k$-space read
\begin{align}
v_k &=   \frac{  \kappa+2\sqrt{2}\, \sin\theta\,\cos k}{\sqrt{4+{\kappa}^2}}, \\
u_k &=  
\frac{ -2\sqrt{2}\,i    \cos\theta\,\sin k}{\sqrt{4+{\kappa}^2}}  
.
\end{align}
From the above expressions, we immediately see that $\Re(u_k v_k^\ast)=0$. Therefore
the purity spectrum $\xi_k$ is equal to $1$ independent of $k$, which yields a completely pure steady state. But the damping matrix shows a richer spectrum with four-fold degenerate relaxation rates with a dispersion
\begin{align}
    \omega_k & = |u_k|^2+|v_k|^2 \nonumber \\
      &= \frac{\big( \kappa+2\sqrt{2}\, \sin\theta\,\cos k \big)^2 +  \big( 2\sqrt{2}\,\cos\theta\,\sin k \big)^2 }{4+{\kappa}^2}.
\end{align}
The dissipative gap of the model ($\Delta_d = {\rm min}\,\omega_k$) is depicted in Fig. \ref{fig:modelI}(a) as a function of 
$\kappa$ and $\theta$. The gap closing in the dissipation spectrum, whereby $\omega_k=0$ for some $k$, takes place when either of the following conditions holds:
\begin{align}
\begin{array}{cc}
     &  \quad (1) \quad   \kappa  =\pm 2\sqrt{2}\, \sin \theta \:, \qquad  \qquad \quad \:\,   \\
     & (2) \quad   \theta = \pm\frac{\pi}{2} \quad {\rm for}\quad |\kappa|\leq 2\sqrt{2} \:.
\end{array}
\end{align}
These gap closing lines are indicated in white in Fig. \ref{fig:modelI}(a).

Since the steady-state of the system is always pure (irrespective of the dissipative couplings parameters), we expect that the topological phase transition, where the winding number changes, must coincide with the dissipative gap closing. By direct evaluation of the winding number $\nu$ as a function of two parameters $\theta$ and $\kappa$ in the jump operators, we see in Fig. \ref{fig:modelI}(b) that indeed the abrupt change in $\nu$ is always accompanied by 
the dissipative gap closing ($\Delta_d=0$). As we see for large enough on-site dissipation $ |\kappa| > 2\sqrt{2}\, \sin \theta $, the steady state of the system is topologically trivial, whereas for $ |\kappa| \leq 2\sqrt{2}\, \sin \theta $,
we have a topologically nontrivial state represented by a nonvanishing $\nu=\pm2$. 
As we have aspired, by introducing the imbalance (which is tuned by $\theta$), we can change the sign of the topological winding number. Putting the result shown in Fig. \ref{fig:modelI}(b) and the Lindblad operator in Eq. \eqref{eq:lindblad-model-I} together, we realize that the winding number's sign is solely determined by the relative sign of excitation and de-excitation coupling terms of the Lindblad operator, i.e. ${\rm sgn}(\tan\theta)$.

\subsection{Model II: With n.n.n. couplings}

\begin{figure}[t]
\centering
\includegraphics[width=0.99\linewidth]{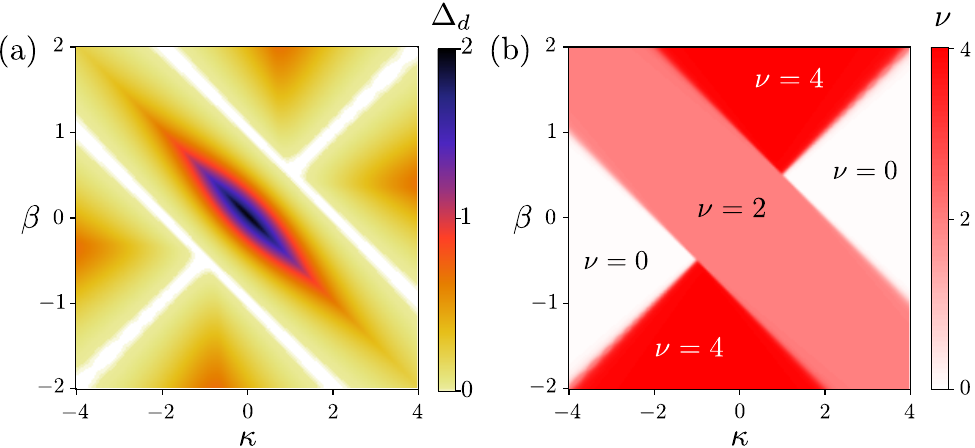}
\caption{(a) Dissipative (damping) gap and (b) the topological winding number
of the Kitaev model, including n.n.n. coupling, as functions of the onsite dissipation parameter $\kappa$ and the relative strength of n.n.n. coupling represented as $\beta$.
Again and similar to the model with only n.n. coupling, the change in the winding number exactly coincides with the dissipative gap closing lines.
For a large enough n.n.n. coupling ($|\beta|>0.5$), we can achieve a higher value of topological invariant, which is $\nu=4$ for this model. 
}
\label{fig:modelII}
\end{figure}

The next extension to the dissipative Kitaev model is achieved by adding n.n.n. couplings denoted by the parameter $\beta$, which leads to the Lindblad operator
\begin{align}
 L^{\rm II}_n  &= \frac{1}{\sqrt{4(1+\beta^2)+{\kappa}^2}} \Big[
 a^{\dagger}_{n+1} + a^{\dagger}_{n-1}
 + a_{n+1} - a_{n-1}
\nonumber \\
&
 + \beta\,(a^{\dagger}_{n+2} + a^{\dagger}_{n-2} + a_{n+2} - a_{n-2} )
 +{\kappa}\,a^{\dagger}_n 
\Big].   \label{eq:lindblad-model-II}
\end{align} 
The model with the above Lindbladian is schematically illustrated in Fig. \ref{fig:lattice}(b),
where we have on-site, n.n., and n.n.n. couplings between the physical sites and auxiliary sites. 
Considering a periodic system, the excitation and de-excitation components of the Fourier transform of the Lindblad operator are
\begin{align}
  &  v_{k} = \frac{\kappa + 2 \cos k +2\beta \cos 2k}{\sqrt{4(1+\beta^2)+{\kappa}^2}}, \\
  &  u_{k} = -2i \frac{ \sin k +\beta \sin 2k }{\sqrt{4(1+\beta^2)+{\kappa}^2}}.
\end{align}
Similar to the previous model, the condition $\Re(u_k v_k^\ast)$ leads to a flat purity spectrum $\xi_k=1$ for all $k$'s. The dispersion relation of the dissipative spectrum is given by 
\begin{align}
    \omega_k
     = \frac{
      \big( \frac{\kappa}{2}+  \cos k +\beta \cos 2k  \big)^2 +  \big(  \sin k +\beta \sin 2k \big)^2 
      }
      {1+\beta^2+\big(\frac{\kappa}{2}\big)^2},
\end{align}
where its gap closing lines are given by 
\begin{align}
\begin{array}{cc}
 &   \kappa = -2(\beta\pm 1) ,\\ 
 &   \qquad\quad  \kappa = 2\beta ~ \quad {\rm for}\quad |\kappa|\geq 1.
\end{array}
\end{align}
To illustrate the dissipative gap, we show its variation with the coupling parameters $\kappa$ and $\beta$ of the Lindblad operator in Fig. \ref{fig:modelII}(a). This result is complemented by the winding number, which is also calculated as a function of $\kappa$ and $\beta$ as shown in Fig. \ref{fig:modelII}(b).
We see that in the absence of n.n.n. couplings ($\beta=0$) only one topologically nontrivial phase with $\nu=2$ is found when the on-site term is bounded to $|\kappa|\leq 2$. This behavior qualitatively persists up to $|\beta|\leq 1/2$, and beyond that, another phase with a higher winding number $\nu=4$ appears. By further increasing $|\beta|$, the range of $\kappa$ over which we have the higher topological invariant inflates as can be seen in Fig. \ref{fig:modelII}(b). 
Similar to the model with imbalanced coupling, the transition between topologically distinct phases is always accompanied by the dissipative gap closing. Since the purity spectrum undergoes no gap closing, the only way to have a topological transition is to close and open the dissipative gap again.

\subsection{Absence of skin effect}
An interesting aspect of the nonequilibrium dynamics governed by the Lindblad equation is their connection to the non-Hermitian models. These connections are beyond the scope of the present work and will be addressed in a separate study. Here, we only briefly discuss the absence of non-Hermitian skin effect in our models even for the imbalanced couplings case present in model I. The skin effect corresponds to the situation where not only the topological edge states but also the bulk states become localized towards the edge. This effect has been considered as a breakdown of the bulk-boundary correspondence, although this issue can be circumvented using the notion of biorthogonality \cite{Kunst2018}. Moreover, there is a topological reason behind the appearance of skin effect, and as it has been shown in Ref. \cite{Sato2020}, it originates from the nontrivial topology of complex energy bands in the presence of point gaps. Now, one way to see if the skin effects exist in a Lindblad dynamic is to check the effective non-Hermitian Hamiltonian associated with the Lindblad operators given by \cite{Bergholtz-RMP}
\begin{equation}
    \tilde{H}_{\rm NH} = H -\frac{i}{2} \sum_{n} L_{n}^\dag L_{n}.
\end{equation}
Since we have focused on $H=0$ throughout the paper, the effective non-Hermitian Hamiltonian becomes anti-Hermitian with a purely imaginary spectrum.
Therefore, the complex spectrum has no point gap which indicates that the skin effect is absent in our models. The absence of the skin effect can be understood intuitively by noting that these models respect the reciprocity even in the presence of imbalanced couplings.

\section{Conclusions and Discussion}
Here, we have provided a generalization of the dissipative Kitaev models, which can easily incorporate higher values of topological invariant and thereby more Majorana bound states as pure Dark states in their stationary long-term behavior. We have found that the inclusion of coupling between the next-nearest-neighbor sites to the bath as well as imbalanced couplings can serve as a tool to engineer topologically distinct phases. An important feature of generalized dissipative models is that they are different from their equilibrium Hamiltonian counterparts. In particular, unlike the hopping strengths in lattice Hamiltonians, we can consider unequal or imbalanced couplings to the bath, which results in a much richer phase diagram.

The simple version of the dissipative Kitaev model involving only couplings between neighboring sites to the bath has been studied in previous theoretical works. This model only shows a constant topological invariant corresponding to the presence of two Majorana bound states at each end. By extending this model with slightly more neighbors involved in dissipative couplings, we have demonstrated that more number of Majorana states can be achieved. Also, we have found that, by tuning the imbalance in nearest-neighbor couplings, we can switch between positive and negative values of topological invariants. Our study is an early attempt in this direction, and many aspects remain to be studied. One intriguing question is about the connections between the nonequilibrium dynamical models based on the Lindblad equation and the non-Hermitian topological systems.

Finally, in light of existing experiments using cold-atom and photonic-based settings to explore dissipative topological phases, we believe the generalized dissipative Kitaev models can be realized in currently accessible experimental setups.

\appendix
\section{Dynamics of correlation operator}\label{app-a}
We first consider the adjoint Lindblad Eq. \eqref{lindblad-adjoint} for number operator $a_q^\dag a_q $, which reads
\begin{align}
    \label{lindblad-adjoint-adaga}
\partial_t (a_q^\dag a_q)=
   \sum_{k} \Big( L_k^\dag  a_q^\dag a_q  L_k - \frac{1}{2}\{ L_k^\dag L_k, a_q^\dag a_q \}\Big).
\end{align}
Now we can easily verify the following anti-commutation relations  
\begin{align}
  \{ a_q,L_k \}&= v_{k} \{a_q,a_k^\dag\} - u_k \{a_q,a_{-k}\} = v_k \delta_{q,k}, \\
  \{ a_q^\dag ,L_k \}&= v_{k} \{a^\dag_q,a_k^\dag\} - u_k \{a^\dag_q,a_{-k}\} = -u_k \delta_{q,-k}, 
\end{align}
which immediately yield the commutation relation
\begin{align}\label{commut-eq-1}
[ a_q^\dag a_q,L_k ] & = a_q^\dag\: \{a_q,L_k \} - \{ a_q^\dag, L_k \} a_q \nonumber \\
&= 
v_q a_q^\dag \:\delta_{q,k} + u_{-q} a_q \:\delta_{-q,k} .
\end{align}
Now using \eqref{commut-eq-1}, the dynamical equation for $a_q^\dag a_q $ can be recast as
\begin{align}
   \partial_t(a_q^\dag a_q)= \frac{1}{2}  \sum_{k} \big[ L_k^\dag L_k, a_q^\dag a_q \big] + 
     v_q L_q^\dag a_q^\dag + u_{-q} L_{-q}^\dag a_q.
\end{align}
We further see that
\begin{align}
    \big[ L_k^\dag L_k, a_q^\dag a_q \big] &= \big[ L_k^\dag L_k, a_q^\dag  \big] a_q +
     a_q^\dag  \big[ L_k^\dag L_k, a_q \big]  \nonumber \\
    & =    L_k^\dag  \big\{L_k, a_q^\dag  \big\} a_q -
    \big\{ L_k^\dag , a_q^\dag  \big\} L_k a_q  \nonumber \\
    & +
     a_q^\dag L_k^\dag \big\{   L_k, a_q \big\}
      -
     a_q^\dag  \big\{  L_k^\dag , a_q \big\} L_k \nonumber \\
     & 
     =   -u_{-q}  L_{-q}^\dag  a_q \: \delta_{-q,k} 
     -v_q^\ast L_q a_q \: \delta_{q,k} \nonumber \\
    & +
    v_q a_q^\dag L_q^\dag \: \delta_{q,k}
     +
     u_{-q}^\ast a_q^\dag  L_{-q} \:  \delta_{-q,k},
\end{align}
using which besides $u_q^\ast \equiv u_{-q}$
and
$v_q^\ast \equiv v_{-q}$,
we finally obtain the following results for the time evolution of $a_q^\dag a_q$:
\begin{align}
\partial_t(a_q^\dag a_q)&=
|v_{q}|^2 - \big( |u_{q}|^2 +|v_{q}|^2\big)a_q^\dag a_q \nonumber \\
&+\Re\big(u_q v_q^\ast\big) 
\big(
a_{-q} a_q
+ a_q^\dag
 a_{-q}^\dag
\big).\label{eq-adag-a}
\end{align}
Then simply apply the anti-commutation between $a_q^\dag $ and $a_q$
and interchanging $q\to -q$ in Eq. \eqref{eq-adag-a}, we arrive at
\begin{align}
\partial_t( a_{-q}a_{-q}^\dag )&=
|u_{q}|^2 - \big( |u_{q}|^2 +|v_{q}|^2\big)a_{-q}a_{-q}^\dag \nonumber \\
&+\Re\big(u_q v_q^\ast\big) 
\big(
a_{-q} a_q
+ a_q^\dag
 a_{-q}^\dag
\big).\label{eq-a-adag}
\end{align}

Now we turn to the anomalous correlations governed by  
\begin{align} \label{lindblad-adjoint-aa}
\partial_t(a_{-q}  a_q)=
   \sum_{k} \Big( L_k^\dag  a_{-q}  a_q  L_k - \frac{1}{2}\{ L_k^\dag L_k, a_{-q}  a_q \}\Big).
\end{align}
Similar to Eq. \eqref{commut-eq-1}, we can write
\begin{align}\label{commut-eq-1p}
[ a_{-q} a_q,L_k ] & = a_{-q}\: \{a_q,L_k \} - \{ a_{-q}, L_k \} a_q \nonumber \\
&= 
v_q a_{-q} \:\delta_{q,k} - v_{-q} a_q \:\delta_{-q,k} ,
\end{align}
and consequently
\begin{align}
    \big[ L_k^\dag L_k, a_{-q}  a_q \big] &= \big[ L_k^\dag L_k,  a_{-q}  \big] a_q +
     a_{-q}  \big[ L_k^\dag L_k, a_q \big]  \nonumber \\
    & =    L_k^\dag  \big\{L_k,  a_{-q}  \big\} a_q -
    \big\{ L_k^\dag , a_{-q}   \big\} L_k a_q  \nonumber \\
    & +
      a_{-q}  L_k^\dag \big\{   L_k, a_q \big\}
      -
      a_{-q}   \big\{  L_k^\dag , a_q \big\} L_k \nonumber \\
     & 
     =   v_{-q}  L_{-q}^\dag  a_q \: \delta_{-q,k} 
     +u_{q}^\ast L_{q} a_q \: \delta_{q,k} \nonumber \\
    & +
    v_q a_{-q} L_q^\dag \: \delta_{q,k}
     +
     u_{-q}^\ast a_{-q}  L_{-q} \:  \delta_{-q,k},
\end{align}
Using the above identities, we can recast Eq.
\eqref{lindblad-adjoint-aa} as the following:
\begin{align}
\partial_t(a_{-q}  a_q)=
   \frac{1}{2} \sum_{k} \big[ L_k^\dag L_k, a_{-q}  a_q \big]
   + 
     v_q L_q^\dag a_{-q} - v_{-q} L_{-q}^\dag a_q ,
\end{align}
and eventually, 
  \begin{align}   
\partial_t(a_{-q}  a_q)&=     
-v_{q} u_{q}^\ast - \big( |u_{q}|^2 +|v_{q}|^2\big)a_{-q} a_q \nonumber \\
&+\Re\big(u_q v_q^\ast\big) 
\big(
a^\dag_q a_q
+ a_{-q}
 a_{-q}^\dag
\big).\label{eq-a-a}
\end{align}
By taking the Hermitian conjugate of the final result for the anomalous term, we also find
\begin{align}
\partial_t(a_{q}^\dag  a_{-q}^\dag)&=     
-v_{q}^\ast u_{q} - \big( |u_{q}|^2 +|v_{q}|^2\big)a_{q}^\dag  a_{-q}^\dag \nonumber \\
&+\Re\big(u_q v_q^\ast\big) 
\big(
a^\dag_q a_q
+ a_{-q}
 a_{-q}^\dag
\big).\label{eq-adag-adag}
\end{align}

Putting Eqs. \eqref{eq-adag-a}, \eqref{eq-a-adag}, \eqref{eq-a-a}, and \eqref{eq-adag-adag} altogether,  we finally acquire the dynamical equation

\begin{widetext}
\begin{align*}
\partial_t \begin{pmatrix}
    a_{q}^\dag a_{q}\\
    a_{-q} a_{-q}^\dag\\
    a_{-q} a_{q}\\
    a_{q}^\dag a_{-q}^\dag
\end{pmatrix}
=-
\begin{pmatrix}
|u_{q}|^2 +|v_{q}|^2 &0& -\Re\big(u_q v_q^\ast\big) &    -\Re\big(u_q v_q^\ast\big)   \\
0& |u_{q}|^2 +|v_{q}|^2 & -\Re\big(u_q v_q^\ast\big) &    -\Re\big(u_q v_q^\ast\big)  \\
-\Re\big(u_q v_q^\ast\big) &    -\Re\big(u_q v_q^\ast\big) & |u_{q}|^2 +|v_{q}|^2 &0 \\
-\Re\big(u_q v_q^\ast\big) &    -\Re\big(u_q v_q^\ast\big) &0 & |u_{q}|^2 +|v_{q}|^2
\end{pmatrix}
\begin{pmatrix}
    a_{q}^\dag a_{q}\\
    a_{-q} a_{-q}^\dag\\
    a_{-q} a_{q}\\
    a_{q}^\dag a_{-q}^\dag
\end{pmatrix}+
\begin{pmatrix}
    |v_q|^2\\
    |u_q|^2\\
    - v_q u_q^\ast\\
    -u_q v_q^\ast
\end{pmatrix},
\end{align*}
\end{widetext}
which in a more compact form can be written as Eq. \eqref{key-dynamics} alongside with definitions given by Eq. \eqref{key-dynamics-details}.

\bibliography{refs}

\end{document}